\documentclass{article}

\usepackage{dsfont,amscd,amsmath,amssymb}

\newcommand{\mq}{\mathbf{q}}
\newcommand{\mpsi}{\boldsymbol{\psi}}
\newcommand{\mbpsi}{\bar{\boldsymbol{\psi}}}
\newcommand{\ml}{\boldsymbol{l}}
\newcommand{\mbl}{\bar{\boldsymbol{l}}}
\newcommand{\mlambda}{\boldsymbol{\lambda}}
\newcommand{\mblambda}{\bar{\boldsymbol{\lambda}}}
\newcommand{\mLambda}{\boldsymbol{\Lambda}}
\newcommand{\mbLambda}{\overline{\boldsymbol{\Lambda}}}
\newcommand{\mT}{\boldsymbol{T}}
\newcommand{\mbT}{\overline{\boldsymbol{T}}}
\newcommand{\mY}{\boldsymbol{Y}}

\newcommand{\p}[1]{(\ref{#1})}

\newcommand{\cF}{{\cal F}}
\newcommand{\cG}{{\cal G}}

\newcommand{\cD}{{\cal D}}
\newcommand{\cE}{{\cal E}}

\newcommand{\bV}{{\overline V}}

\newcommand{\bZ}{{\overline Z}}

\newcommand{\bD}{{\overline D}}

\newcommand{\bQ}{{\overline Q}}
\newcommand{\bS}{{\overline S}}
\newcommand{\bK}{{\overline K}}
\newcommand{\bLambda}{{\overline \Lambda}{}}

\newcommand{\btheta}{{\bar\theta}}
\newcommand{\bpsi}{{\bar\psi}{}}
\newcommand{\blambda}{{\bar\lambda}{}}

\newcommand{\bq}{{\bar q}}
\newcommand{\bnabla}{{\overline{\nabla}}}

\newcommand{\be}{\begin{equation}}
\newcommand{\ee}{\end{equation}}
\newcommand{\bea}{\begin{eqnarray}}
\newcommand{\eea}{\end{eqnarray}}

\newcommand{\ba}{\begin{array}} \newcommand{\ea}{\end{array}}

\def\im{{\rm i}}

\newcommand{\nn}{\nonumber}

\usepackage{amscd,amsmath,amssymb}

\def\theequation{\arabic{section}.\arabic{equation}}

\begin{document}
\thispagestyle{empty}
\begin{flushright}
\end{flushright}

\vspace{3cm}

\begin{center}
{\Large\bf Supermembrane in $D=5$: component action}
\end{center}
\vspace{1cm}

\begin{center}
{\large\bf S.~Bellucci${}^a$, N.~Kozyrev${}^b$, S.~Krivonos${}^{b}$,
 A.~Yeranyan${}^{c,a,d}$ }
\end{center}

\begin{center}
${}^a$ {\it
INFN-Laboratori Nazionali di Frascati,
Via E. Fermi 40, 00044 Frascati, Italy} \vspace{0.2cm}

${}^b$ {\it
Bogoliubov  Laboratory of Theoretical Physics, Joint Institute for Nuclear Research,
141980 Dubna, Russia} \vspace{0.2cm}

${}^c$ {\it
Museo Storico della Fisica e Centro Studi e Ricerche " Enrico Fermi",
Via Panisperna 89A, 00184 Roma, Italy} \vspace{0.2cm}

${}^c$ {\it
Department of Physics, Yerevan State University,
Alex Manoogian St. 1, Yerevan, 0025, Armenia} \vspace{0.2cm}

\end{center}
\vspace{2cm}

\begin{abstract}\noindent
Based on the connection between partial breaking of global
supersymmetry, coset approach, which realized the given pattern of
supersymmetry breaking, and the Nambu-Goto actions for the
extended objects, we have constructed on-shell component action
for $N=1, D=5$ supermembrane and its dual cousins. We demonstrate
that the proper choice of the components and the use of the
covariant (with respect to broken supersymmetry) derivatives
drastically simplify the action: it can be represented as a sum of
four terms each having an explicit geometric meaning.
\end{abstract}

\newpage
\setcounter{page}{1}
\setcounter{equation}{0}
\section{Introduction}
The characteristic feature of the theories with a partial breaking
of the global supersymmetries is the appearance of the Goldstone
fermionic fields, associated with the broken supertranslations, as
the components of Goldstone supermultiplets of unbroken
supersymmetry. The natural description of such theories is
achieved within the coset approach \cite{NR, NR1}. The usefulness
of the coset approach in the applications to the theories with
partial breaking of the supersymmetry have been demonstrated by
many authors [3-20]. The presence of the unbroken supersymmetry
makes quite reasonable the idea to choose the corresponding
superfields as the basic ones and many interesting superspace
actions describing different patterns of supersymmetry breaking
have been constructed in such a way \cite{BG1,RT,8,IK1}. However,
the standard methods of coset approach fail to construct the
superfield action, because the superspace Lagrangian is weakly
invariant with respect to supersymmetry - it is shifted by the
full space-time or spinor derivatives under broken/unbroken
supersymmetry transformations. Another, rather technical
difficulty is to obtain the component action from the superspace
one, which is written in terms of the superfields subjected to
highly nonlinear constraints. Finally, in some cases the
covariantization of the irreducibility constraints with respect to
the broken supersymmetry is not evident, if at all possible. For
example, it has been demonstrated in \cite{BG1} that such
constraints for the vector supermultiplet can be covariantized
only together with the equations of motion.

It turned out that one can gain more information about component
off-shell actions if attention is shifted to the broken
supersymmetry. It was demonstrated in \cite{BKS1} that with a
suitable choice of the parametrization of the coset, the
$\theta$-coordinates of unbroken supersymmetry and the physical
bosonic components do not transform under broken supersymmetry.
Moreover, the physical fermions transform as the Goldstino of the
Volkov-Akulov model \cite{VA} with respect to broken
supersymmetry. Therefore, the physical fermions can enter the
component on-shell action only i) through the determinant of the
fermionic vielbein (to compensate the variation of the volume $d^d
x$), ii) through the covariant space-time derivatives, or iii)
through the Wess-Zumino term, if it exists. The first two
ingredients can be easily constructed within the coset method,
while the Wess-Zumino can be also constructed from Cartan forms
following the recipe of ref.\cite{HM}. As a result, we will have
the Ansatz for the action with several constant parameters, which
have to be fixed by the invariance with respect to unbroken
supersymmetry. The pleasant feature of such an approach is that
the fermions are ``hidden'' inside covariant derivatives and
determinant of the vielbein, making the whole action short, with
the explicit geometric meaning of each term. In the present paper
we apply this procedure to construct the action of $N=1, D=5$
supermembrane and its dual cousins.

\setcounter{equation}{0}
\section{Supermembrane in $D=5$ space-time}
In accordance with the general consideration presented in
\cite{BKS1}, to construct the component action for the
supermembrane in $D=5$ one has to carry out the following steps:
\begin{itemize}
\item Choosing the proper parametrization of the  coset space element corresponding to the given pattern of the supersymmetry breaking;
constructing the Cartan forms and finding the covariant
derivatives,
\item Imposing the kinematical and dynamical constraints,
\item Finding the bosonic limit of the action and then generalizing it to the full supersymmetric case,
\item Fixing the arbitrary constants in the supersymmetric action by imposing the invariance
with respect to  unbroken supersymmetry.
\end{itemize}
Let us perform this programme.

\subsection{Coset space}
In the present case we are dealing with the spontaneous breaking
of $N=1, D=5$ Poincar\'{e} supersymmetry down to $N=2, d=3$ one.
From the $d=3$ standpoint the $N=1, D=5$ supersymmetry algebra is
a central-charges extended $N=4$ Poincar\'{e} superalgebra with
the following basic anticommutation relations: \be\label{basicAL}
\left\{ Q_{a} , \bQ_{b}  \right\} =2P_{ab}, \; \left\{ S_a, \bS_b
\right\}=2P_{ab},  \; \left\{ Q_{a}, S_{b} \right\}
=2\epsilon_{ab} Z, \; \left\{ \bQ_{a}, \bS_{b} \right\}
=2\epsilon_{ab} \bZ. \ee The $d=3$ translations generator $P_{ab}$
and the central charge generators $Z,\bZ$ form $D=5$ translation
generators. We will also split the generators of $D=5$ Lorentz
algebra $so(1,4)$ into $d=3$ Lorentz algebra generators $M_{ab}$,
the generators $K_{ab}$ and $\bK_{ab}$ belonging to the coset
$SO(1,4)/SO(1,2)\times U(1)$ and the $U(1)$ generator $J$. The
full set of commutation relations can be found in the Appendix A,
\p{algebra}.

Keeping $d=3$ Lorentz and, commuting with it, $U(1)$ subgroups of
$D=5$ Lorentz group $SO(1,4)$ linearly realized, we will choose
the coset element as \be\label{a_coset} g = e^{\im
x^{ab}P_{ab}}e^{\theta^a Q_a + \bar\theta^a \bQ_a}e^{\im(\mq
Z+\bar\mq \bZ)}e^{\mpsi^a S_a + \mbpsi^a \bS_a} e^{\im
(\mLambda^{ab}K_{ab}+ \mbLambda^{ab}\bK_{ab})}. \ee Here,
$\left\{x^{ab}, \theta^a, \bar\theta^a\right\}$ are $N=2, d=3$
superspace coordinates, while the remaining coset parameters are
$N=2$ Goldstone superfields. The whole $N=1, D=5$ super
Poincar\'{e} group can be realized in this coset by the left
acting on \p{a_coset} of the different elements of the supergroup.
The resulting transformation properties of the coordinates and
superfields with respect to unbroken and broken supersymmetries
are presented in \p{susy1}, \p{susy2}. The results of a pure
technical calculation of the corresponding Cartan forms,
semi-covariant derivatives and their algebra are summarized in the
Appendix A, \p{CF}, \p{nabla}, \p{deralg}.

\subsection{Kinematical constraints and equations of motion}
In accordance with the general theorem (Inverse Higgs phenomenon)
formulated in \cite{ih}, in order to reduce the number of
independent superfields one has to impose the constraints
\be\label{IH} \Omega_Z =0 \quad \Rightarrow \quad \left\{
\begin{array}{l}
\nabla_{ab}\mq =- 2\im \frac{(1+\ml \cdot \mbl) \ml_{ab} -\ml^2 {\mbl}_{ab}}{(1+\ml\cdot\mbl)^2-\ml^2{\mbl}{}^2},\\
\nabla_a \mq = -2\im \mpsi_a,\quad \bar\nabla_a \mq =0,
\end{array} \right. \quad
{\overline\Omega}_Z =0 \quad \Rightarrow \quad \left\{
\begin{array}{l}
\nabla_{ab}{\bar\mq} = 2\im \frac{(1+\ml \cdot \mbl) \mbl_{ab} -\mbl^2 \ml_{ab}}{(1+\ml\cdot\mbl)^2-\ml^2{\mbl^2}},\\
\bar\nabla_a {\bar\mq} = -2\im \bar\mpsi_a,\quad \nabla_a {\bar\mq} =0.
\end{array} \right.
\ee Here, to simplify the expressions, we have passed to the some
variant of the stereographic parametrization of the coset
$SO(1,4)/SO(1,2)\times U(1)$ \be\label{l} \ml_{ab}=\left(
\frac{\tanh\sqrt{\boldsymbol{Y}}}{\sqrt{\boldsymbol{Y}}}\right)_{ab}^{cd}\;\mLambda_{cd},\quad
\mbl_{ab}=\left(
\frac{\tanh\sqrt{\boldsymbol{Y}}}{\sqrt{\boldsymbol{Y}}}\right)_{ab}^{cd}\;\mbLambda_{cd}.
\ee The equations \p{IH} allow us to express the superfields
$\mLambda_{ab}, \mbLambda_{ab}$ and $\mpsi^a, \mbpsi^a$ through
covariant derivatives of $\mq(x,\theta,\bar\theta)$ and
${\bar\mq}(x,\theta,\bar\theta)$. Thus, the bosonic superfields
$\mq(x,\theta,\bar\theta), {\bar\mq}(x,\theta,\bar\theta)$ are the
only essential Goldstone superfields  needed for this case of the
partial breaking of the global supersymmetry. The constraints
\p{IH} are covariant under all symmetries, they do not imply any
dynamics and leave $\mq(x,\theta,\bar\theta)$ and
${\bar\mq}(x,\theta,\bar\theta)$ off shell.

Within the coset approach we may also to write the covariant superfield equations of motion. It was shown in
[11-17] that this can be achieved by imposing the proper constraint on the
 Cartan forms for broken supersymmetry. In the present case these constraints read
\bea\label{eomCF}
&& \left. \Omega_S \right|=0\; \Rightarrow \; (a)\; \nabla_a \mpsi_b = 0, \quad (b)\; \bar\nabla_b \mpsi^a=- \im \mLambda_b{}^c
\left( \frac{\tan 2 \sqrt{ \overline{\boldsymbol{T}}}}{ \sqrt{ \overline{\boldsymbol{T}}}}\right)_c^a \equiv
-\im \mlambda_b^a \nn \\
&& \left. {\overline\Omega}_S \right|=0\; \Rightarrow \; (a)\;
\bar\nabla_a \mbpsi_b = 0, \quad (b)\; \nabla_b \mbpsi^a= \im
\mbLambda_b{}^c \left( \frac{\tan 2 \sqrt{ {\boldsymbol{T}}}}{
\sqrt{ {\boldsymbol{T}}}}\right)_c^a \equiv \im \mblambda_b^a,
\eea where $|$ means the $d\theta$-projection of the forms. These
constraints are closely related with constraints of the
super-embedding approach \cite{Dima}.

To conclude this Subsection let us make a few comments:
\begin{itemize}
\item The easiest way to check that the equations \p{IH}, \p{eomCF} put the theory on-shell  is to consider these equations in the linearized form
    \bea
    && \partial_{ab} \mq =-2\im \mLambda_{ab} \; (a), \quad D_a \mq =-2\im \mpsi_a\; (b),\quad \bD_a \mq=0 \; (c),\label{lin1} \\
    && D_a \mpsi_b =0 \; (a), \quad \bD_b \mpsi^a =- 2 \im \mLambda_b^a\; (b). \label{lin2}
    \eea
    Acting on eq.(\ref{lin1}b) by $\bD_b$ and using the eq.(\ref{lin1}c) and the algebra of spinor derivatives \p{flatCD} we immediately conclude
     that eq.(\ref{lin2}b) follows from \p{lin1}. In addition, the eq.(\ref{lin2}a)
    means that the auxiliary component of the superfield $\mq$ is zero and, therefore, our system is on-shell
    \be\label{lin3}
    D_a \mpsi_b =0\; \Rightarrow \; D^2 \mq =0\; \Rightarrow \partial_{ab} D^b \mq =0\;\Rightarrow \Box \mq=0.
    \ee
\item It turns out that the variables $\{\mlambda_a^b, \mblambda_a^b\}$ defined in \p{eomCF}, are more suitable then
$\{\ml_{ab}, \mbl_{ab}\}$ \p{l} one. Using the algebra of
covariant derivatives \p{deralg} it is easy to find the following
relations from \p{IH} and \p{eomCF} \be\label{qlambda}
\nabla_{ab}\mq =-\im \frac{ \mlambda_{ab}-\frac{1}{2} \mlambda^2
\mblambda_{ab}}{1-\frac{1}{4} \mlambda^2 \mblambda^2},\quad
\nabla_{ab}\bar\mq =\im \frac{ \mblambda_{ab}-\frac{1}{2}
\mblambda^2 \mlambda_{ab}}{1-\frac{1}{4} \mlambda^2 \mblambda^2}.
\ee These equations play the same role as those in \p{IH},
relating the superfields $\{\mlambda_{ab}, \mblambda_{ab}\}$ (and,
therefore, the superfields $\{\mLambda_{ab}, \mbLambda_{ab}\}$)
with the space-time derivatives of the superfields $\{\mq,
\bar\mq\}$.
\end{itemize}
\subsection{Bosonic part}
In what follows we will mainly deal with the component approach.
So, let us define the components of our superfields as
\be\label{defcomp} q=\mq|_{\theta=0}, \quad \psi_a
=\mpsi_a|_{\theta=0},\quad \lambda_{ab}
=\mlambda_{ab}|_{\theta=0},\quad
\Lambda_{ab}=\mLambda_{ab}|_{\theta=0}. \ee The basic idea of the
approach outlined in \cite{BKS1} is to write the candidate for
supersymmetric action as a proper supersymmetrization of the
bosonic action. So, the crucial step is to construct the bosonic
action. Keeping in the mind that the system of equations \p{IH},
\p{eomCF} is invariant with respect to {\it all} $N=1, D=5$ super
Poincar\'{e} group, we have to conclude that its bosonic
sub-sector has to be invariant under the bosonic part of the
super-Poincar\'{e} group, i.e. under $ISO(1,4)$ transformations.
This information is enough to construct the bosonic action.

There are, at least, three equivalent ways to construct the bosonic action.
\subsubsection{Bosonic coset}
The simplest, straightforward way to construct the bosonic action is to consider the bosonic coset,
i.e. the coset \p{a_coset} with discarded $\theta$'s and all fermions
\be\label{bos_coset}
g_{bos} = e^{\im x^{ab}P_{ab}}e^{\im( q Z+\bar{q} \bZ)} e^{\im  (\Lambda^{ab}K_{ab}+ \bLambda^{ab}\bK_{ab})}.
\ee
Clearly, the corresponding bosonic Cartan forms can be easily extracted from their superfields version \p{CF}. The bosonic
version of the constraints \p{IH} will result in the relations
\be\label{boIH}
\partial_{ab} q =- 2\im \frac{(1+l \cdot \bar{l}) l_{ab} -l^2 {\bar l}_{ab}}{(1+l\cdot\bar{l})^2-l^2{\bar l}{}^2},\quad
\partial_{ab}{\bar q} = 2\im \frac{(1+l \cdot \bar{l}) \bar{l}_{ab} -{\bar l}^2 l_{ab}}{(1+l\cdot\bar{l})^2-l^2{\bar{l}^2}},
\ee
while the bosonic vielbein $\mathcal{B}_{ab}{}^{cd}$
\be\label{bos_e_def}
\left(\Omega_P^{bos}\right) = dx^{ab} \mathcal{B}_{ab}{}^{cd} P_{cd}
\ee
acquires the form
\be
 \mathcal{B}_{ab}^{cd}=\delta_{a}^{(c} \delta_{b}^{d)}-\frac{2}{(1+l\cdot \bar{l})^2-l^2\, \bar{l}^2}\left[(1+l\cdot \bar{l})\,\left(\bar{l}^{cd} l_{ab}+l^{cd} \bar{l}_{ab}\right)- \bar{l}^2\, l^{cd} l_{ab}- l^2\, \bar{l}^{cd} \bar{l}_{ab}\right]\nn,
\ee
Therefore, the simplest invariant bosonic action reads
\be\label{bos_action1}
S_{bos}=\int d^3x \det B =
 \int d^3x \frac{(1-l\cdot \bar{l})^2-l^2\, \bar{l}^2}{(1+l\cdot \bar{l})^2-l^2\, \bar{l}^2},
\ee or in terms of $\{q, {\bar q}\}$ \be\label{bos_action2}
S_{bos}=\int d^3x \sqrt{\left(1-\partial_{ab}q \,
\partial^{ab}\bar q\right)^2-\left(\partial_{ab}q\,
\partial^{ab}q\right)\,\left(\partial_{cd}\bar q
\,\partial^{cd}\bar q\right)}. \ee The latter is the static gauge
Nambu-Goto action for the membrane in D=5.
\subsubsection{Direct construction}
Another way to derive the bosonic action is to use automorphism
transformation laws. These laws \p{auto} in the bosonic limit have
the form \be\label{d3N2aut1} \delta x^{ab} = 2\im \left( \bar a
^{ab} q - a^{ab} \bar q   \right), \ \delta q =-2\im (ax), \
\delta\bar q =2\im (\bar a x). \ee The active form of these
transformations reads \be\label{d3N2aut2} \delta^\star q = -2\im
(ax)-2\im \partial_{ab} q \left(   \bar a ^{ab} q - a^{ab} \bar q
\right), \ \delta^\star \bq = 2\im (\bar ax)-2\im \partial_{ab}
\bar q \left(   \bar a ^{ab} q - a^{ab} \bar q  \right). \ee Due
to translations, $U(1)$-rotations  and $d=3$ Lorentz invariance,
the action may depend only on scalars $\xi$ and $\left(\eta \bar
\eta\right)$, where \be \xi = \partial_{ab}q \;\partial^{ab}\bar
q,\quad \eta =
\partial_{ab}q\; \partial^{ab}q,\; \bar \eta= \partial_{ab}{\bar
q}\; \partial^{ab}{\bar q} . \ee Their transformation laws can be
easily found to be \bea\label{d3N2aut3}
\delta^\star \xi &=& 2\im (\bar a \partial q) -2\im (a\partial \bar q) -2\im (\bar a^{ab}q  - a^{ab}\bar q)\partial_{ab}\xi -2\im (\bar a \partial q)\xi + 2\im (a \partial \bar q)\xi -2\im (\bar a \partial \bar q)\eta + 2\im (a\partial q)\bar\eta,\nn  \\
\delta^\star (\eta\bar\eta) &=& 4\im (\bar a \partial \bar q)\eta -4\im (a\partial q)\bar\eta -2\im (\bar a^{kl} q - a^{kl} \bar q)\partial_{kl}(\eta\bar\eta)-4\im (\bar a \partial q) \eta\bar\eta + 4\im (a\partial \bar q)\eta\bar\eta + \nn \\
&& 4\im (a\partial q)\xi\bar\eta -4\im (\bar a \partial \bar
q)\xi\eta. \eea Therefore, the variation of the arbitrary function
$F(\xi,\eta\bar\eta)$ reads \bea\label{d3N2aut4} \frac{1}{2\im}
\delta^\star F &=& \left[ (a\partial q)\bar\eta - (\bar a \partial
\bar q) \eta \right]\left( F_\xi + 2(\xi-1)F_{(\eta\bar\eta)}
\right) +  \left[ (\bar a \partial q) - (a\partial \bar q)
\right]\left(F+ (1-\xi)F_\xi -2\eta\bar\eta F_{(\eta\bar\eta)}
\right) -
\nn \\
&&\partial_{ab}\left[ \left(q {\bar a}^{ab} - {\bar q}
a^{ab}\right) F \right]. \eea Thus, to achieve the invariance of
the action one has impose the following restrictions on the
function $F$: \be\label{d3N2aut5} F_\xi +2(\xi
-1)F_{(\eta\bar\eta)} =0, \quad F+F_\xi (1-\xi) -2(\eta\bar\eta)
F_{(\eta\bar\eta)} =0, \ee with the evident solution \be F=
\sqrt{(1-\xi)^2 -\eta\bar\eta}. \ee Therefore, the invariant
action has the form
$$
S = \int d^3 x \sqrt{(1-\partial_{ab} q \partial^{ab}\bar q)^2 -(\partial_{ab} q \partial^{ab} q)(\partial_{kl}\bar q \partial^{kl}\bar q)} ,
$$
and thus, it coincides with the previously constructed one in
\p{bos_action2}, as it should be. Finally, one should note that
the trivial action \be\label{bosS0} S_0 = \alpha \int d^3 x,
\qquad \alpha=const \ee is also invariant under $ISO(1,4)$
transformations.
\subsubsection{Using the equations of motion}
This way is more involved, thus we just sketch main steps. The idea is to find the bosonic equations of motion for $\{q, {\bar q}\}$,
which follow from \p{IH},\p{eomCF}. These equations will explicitly contain  $\{ \lambda_{ab}, \blambda_{ab}\}$, which have to be expressed
through $\{\partial_{ab} q, \partial_{ab}{\bar q}\}$ from the bosonic version
of the equations \p{qlambda}
\be\label{boqlambda}
\partial_{ab} q =-\im \frac{ \lambda_{ab}-\frac{1}{2} \lambda^2 \blambda_{ab}}{1-\frac{1}{4} \lambda^2 \blambda^2},\quad \partial_{ab}\bar{q} =\im \frac{ \blambda_{ab}-\frac{1}{2} \blambda^2 \lambda_{ab}}{1-\frac{1}{4} \lambda^2 \blambda^2}.
\ee
Having at hands the equations of motion one may reconstruct the bosonic action, which of course, will again coincide with \p{bos_action2}.

Clearly, in the present case the two previously discussed  ways
are simpler. Nevertheless, in the cases where some of the physical
bosonic components have no Goldstone fields interpretation, this
way is rather efficient, if not the simplest ones (see e.g.
\cite{BIK22}).

\subsection{Adding supersymmetry}
Now, we have at hands all ingredients to construct the full component action for the membrane which will be invariant
under both, broken $S$ and unbroken $Q$ supersymmetries. In our approach, we are starting with the broken supersymmetry.

\subsubsection{Broken $S$ supersymmetry}
In our parametrization of the coset \p{a_coset} the superspace coordinates $\{\theta,\bar\theta\}$ do not transform under
$S$ supersymmetry. Therefore, each component of our superfields transforms independently and from \p{susy2} one may find
that
\be\label{Str}
\delta x^{ab}= \im \left(\varepsilon^{(a}\bpsi^{b)}+\bar\varepsilon^{(a}\psi^{b)}\right),\quad
\delta q=0,\;\delta \bar{q}=0,\quad \delta\psi^a=\varepsilon^a,\;\delta\bpsi^a=\bar\varepsilon^a\; .
\ee
Then, one may easily check that the $\theta=0$ projections of the covariant differential $\triangle x^{ab}$ \p{Dx}
\be\label{Sdx}
\hat{\triangle} x^{ab} \equiv \triangle x^{ab}|_{\theta=0} = dx^{ab} -\im \left(
\psi^{(a} d\bpsi^{b)} + \bpsi^{(a} d\psi^{b)}\right)\equiv \cE^{ab}_{cd}\; dx^{cd},
\ee
as well as the covariant derivatives constructed from them
\be\label{SD}
\cD_{ab} = \left( \cE^{-1} \right)_{ab}^{cd}\; \partial_{cd}
\ee
are also invariant under $S$ supersymmetry. From all these it immediately follows that the action possessing the proper bosonic limit
\p{bos_action2} and invariant under broken supersymmetry reads
\be\label{Saction}
S_1 = \int d^3 x\; \det \cE \sqrt{(1-\cD_{ab} q \cD^{ab}\bar q)^2 -(\cD_{ab} q \cD^{ab} q)(\cD_{cd}\bar q \cD^{cd}\bar q)} .
\ee
The action $S_1$  reproduces the kinetic terms for the bosonic and fermionic components
\be\label{Saction_lin}
S_1 = \int d^3 x\;\left[ -\im \left( \psi^a \partial_{ab} \bpsi^b + \bpsi^a \partial_{ab} \psi^b \right) -
\partial_{ab}q \partial^{ab}{\bar q} + \ldots\right],
\ee but the coefficient between these kinetic terms is strictly
fixed. This could be not enough to maintain $Q$ supersymmetry. So,
one has to add to the action $S_1$ the purely fermionic action
$S_2$ \be\label{Saction2} S_2 =\int d^3 x\; \det \cE , \ee which
is trivially invariant under $S$ supersymmetry. Finally, to have a
proper limit
$$
S_{q\rightarrow 0, \psi \rightarrow 0} =0,
$$
one has to involve into the game the trivial action $S_0$
\be\label{Saction0} S_0 = \int d^3 x. \ee Thus, our anzatz for the
supersymmetric action acquires the form \bea\label{Saction_gen}
S&=& \left( 1 + \alpha\right) S_0 - S_1 - \alpha S_2=\nn \\
&&\left( 1 + \alpha\right)\int d^3 x - \int d^3 x \det \cE \left(
\alpha+  \sqrt{(1-\cD_{ab} q \cD^{ab}\bar q)^2 -(\cD_{ab} q
\cD^{ab} q)(\cD_{cd}\bar q \cD^{cd}\bar q)}\right)  , \eea where
$\alpha$ is a constant that has to be defined.

In the cases previously considered within the present approach
\cite{BKS1,BKKS1}, the Ansatz, similar to \p{Saction_gen}, was
completely enough to maintain the second, unbroken supersymmetry.
The careful analysis shows that in the present case there is one
additional, Wess-Zumino term which has to be taken into account
\be\label{WZ} S_{WZ}= \im  \int d^3 x \;\det \cE \;\left( \psi^m
\cD_{ab} \bpsi_m - \bpsi^m \cD_{ab} \psi_m \right) \cD^{ac}\; q\;
\cD_c{}^b \; {\bar q} . \ee The variation of the $S_{WZ}$ under
$S$ supersymmetry reads (only the variations of $\psi, \bpsi$
without derivatives play a role) \be\label{inv1} \delta S_{WZ}
=\im  \int d^3 x \;\det \cE \;\left( \varepsilon^m \cD_{ab}
\bpsi_m - \bar\varepsilon^m \cD_{ab} \psi_m \right) \cD^{ac}\; q\;
\cD_c{}^b \; {\bar q}. \ee The simplest way to check that $\delta
S_{WZ}=0$ is to pass to the $d=3$ vector notations\footnote{Our
conventions to pass to/from vector indices are summarized in the
Appendix A, \p{44}.}. Then we have \bea\label{inv2}
\delta S_{WZ} &\sim &  \int d^3 x \;\det \cE \;\epsilon^{IJK}\left( \varepsilon^m \cD_I \bpsi_m - \bar\varepsilon^m \cD_I \psi_m \right) \cD_J\; q\; \cD_K \; {\bar q} \sim \nn \\
&& \int d^3 x \;\det \cE \;\det\cE^{-1} \epsilon^{IJK}\left( \varepsilon^m \partial_I \bpsi_m - \bar\varepsilon^m \partial_I \psi_m \right) \partial_J\; q\; \partial_K \; {\bar q} \sim \nn \\
&&\int d^3 x \;\partial_I\left[ \epsilon^{IJK}\left( \varepsilon^m
\bpsi_m - \bar\varepsilon^m \psi_m \right) \partial_J\; q\;
\partial_K \; {\bar q}\right] =0. \eea Thus, the action $S_{WZ}$
\p{WZ} is invariant under $S$ supersymmetry and our Ansatz for the
membrane action extended to be \be\label{ACTION} S= \left( 1 +
\alpha\right) S_0 - S_1 - \alpha S_2+\beta S_{WZ}. \ee

Let us stress, that after imposing broken supersymmetry, our
component action \p{ACTION} is fixed up to two constants $\alpha$
and $\beta$. No other terms or structures are admissible! Funny
enough, the role of the unbroken supersymmetry is just to fix
these constants.

\subsubsection{Unbroken supersymmetry}
To maintain the unbroken supersymmetry, firstly, one has to find the transformation properties of the components. Using the
transformations of the super-space coordinates \p{susy1}
$$
\delta \theta^a = \epsilon^a, \; \delta \bar \theta^a = \bar\epsilon^a, \quad \delta x^{ab} = \im (\epsilon^{(a} \bar \theta^{b)} + \bar\epsilon^{(a} \theta^{b)}),
$$
one may easily find the transformations of the needed ingredients
(we will explicitly present only $\epsilon$-part of the transformations):
\bea
\delta\psi_a &=& -\epsilon^b \left.\left(D_b \mpsi_a\right)\right|_{\theta=0} = \epsilon^b \psi^m\blambda_b^n \partial_{mn}\psi_a,\nn\\
\delta\cD_{ab}\psi_c&=&-\epsilon^d \left.\left( D_d \nabla_{ab}\mpsi_c\right)\right|_{\theta=0} =
2 \epsilon^d \cD_{ab}\psi^m \blambda_d^n \cD_{mn}\psi_b+\epsilon^d \psi^m\blambda_d^n\partial_{mn}\cD_{ab}\psi_c, \nn\\
\delta\cD_{ab}q&=&-\epsilon^d \left.\left( D_d
\nabla_{ab}q\right)\right|_{\theta=0} = 2\epsilon^d \cD_{ab}\psi^m
\blambda_d^n\cD_{mn} q +2\im \epsilon^d \cD_{ab} \psi_d+
\epsilon^d\psi^m \blambda_d^n \partial_{mn}\cD_{ab} q,
\label{Qsusytr} \eea and, as a consequence, \be\label{SEtr} \delta
\det\cE= \partial_{mn}\left[ \epsilon^d \psi^m \blambda_d{}^n
\det\cE\right] - 2 \epsilon^d \blambda_d{}^n \cD_{mn}\psi^n
\det\cE . \ee

To fix the parameter $\alpha$ one may consider just the kinetic
terms in the action \p{ACTION} \be\label{Skin} S_{kin} = \int d^3
x \left[ -\im \left(\alpha+1\right) \left( \psi^a \partial_{ab}
\bpsi^b+ \bpsi^a \partial_{ab}\psi^b\right) +\partial_{ab} q
\partial^{ab}{\bar q}\right], \ee which has to be invariant under
linearized transformations \p{Qsusytr} \be\label{linQ} \delta
\bpsi_a =- \im \epsilon^b\blambda_{ba}\simeq-\epsilon^b
\partial_{ba}{\bar q}, \quad \delta\partial_{ab}q = 2\im
\epsilon^d \partial_{ab}\psi_d. \ee Varying the integrand in
\p{Skin} and integrating by parts, we will get \be \delta S_{kin}
= \int d^3 x\left[ 2\im (\alpha+1) \epsilon^c \psi^a\partial_{ab}
\partial_c{}^b {\bar q} - 2\im \epsilon^d \psi_d \Box {\bar q}
\right] = \int d^3 x\left[ \im (\alpha+1) \epsilon^d \psi_d \Box
{\bar q} - 2\im \epsilon^d \psi_d \Box {\bar q} \right]. \ee
Therefore, we have to fix \be\label{alpha} \alpha=1. \ee

Unfortunately, the fixation of the last parameter $\beta$ is more involved. Using the transformation properties\p{Qsusytr}
one may find
\bea
\delta \cF &=& 2\left( \epsilon^c \blambda_c^n \cD_{ab} \psi^m \cD_{nm} q+\im \epsilon^c \cD_{ab}\psi_c\right) \frac{\partial \cF}{\partial \cD_{ab}q}+ 2 \epsilon^c \blambda_c^n \cD_{ab}\psi^m \cD_{mn}{\bar q}\frac{\partial \cF}{\partial \cD_{ab}{\bar q}}+\nn \\
&& \epsilon^c \blambda_c^n \psi^m \partial_{mn} \cF, \eea where
\be\label{F} \cF \equiv \sqrt{(1-\cD_{ab} q \cD^{ab}\bar q)^2
-(\cD_{ab} q \cD^{ab} q)(\cD_{cd}\bar q \cD^{cd}\bar q)}. \ee To
avoid the appearance of the square roots, it is proved to be more
convenient to use the following equalities \be \frac{\partial
\cF}{\partial \cD_{ab}q} =-\im \frac{\blambda^{ab}+\frac{1}{2}
\blambda^2 \lambda^{ab}}{1-\frac{1}{4}\lambda^2 \blambda^2}, \quad
\frac{\partial \cF}{\partial \cD_{ab}{\bar q}} =\im
\frac{\lambda^{ab}+\frac{1}{2} \lambda^2
\blambda^{ab}}{1-\frac{1}{4}\lambda^2 \blambda^2}. \ee After some
straightforward calculations  we get \be \delta \left[- \det\cE
\left(1+\cF\right)\right]= 2 \im \epsilon^c \det\cE \left(
\cD_{ab}\psi_c \cD^{ab}{\bar q} -
2\cD_{am}\psi^m \cD_c^a {\bar q}\right) - 
2 \epsilon^c \blambda_{cm} \cD_{ab}\psi^m  \cD^{ad}q \cD_d^b {\bar
q} \det \cE . \label{deltaF} \ee Similarly, one may find the
variation of the integrand of the action $S_{WZ}$ (up to surface
terms disappearing after integration over $d^3x$) \be \delta {\cal
L}_{WZ}= -2 \beta \det\cE \left[ \left( \psi^k \cD_{ab}\bpsi_k
-\bpsi^k \cD_{ab}\psi_k\right) \epsilon^c \cD^{ad}\psi_c \cD_d^b
{\bar q} -\epsilon^c \blambda_{cm} \cD_{ab}\psi^m \cD^{ad} q
\cD_d^b{\bar q}\right]. \label{deltaWZ} \ee Now, it is a matter of
quite lengthly, but again straightforward calculations, to check
that the sum of variations \p{deltaF} and \p{deltaWZ} is a surface
term if \be\label{beta} \beta=1 . \ee Thus, we conclude that the
action of the supermembrane in $D=5$, which is invariant with
respect unbroken and broken supersymmetries, has the form \bea S
&=& 2\int d^3 x-
\int d^3 x \det \cE \left( 1+  \sqrt{(1-\cD_{ab} q \cD^{ab}\bar q)^2 -(\cD_{ab} q \cD^{ab} q)(\cD_{cd}\bar q \cD^{cd}\bar q)}\right)+ \nn \\
&& \im  \int d^3 x \;\det \cE \;\left( \psi^m \cD_{ab} \bpsi_m - \bpsi^m \cD_{ab} \psi_m \right) \cD^{ac}\; q\; \cD_c{}^b \; {\bar q} . \label{finAction}
\eea

\setcounter{equation}{0}
\section{Dualization of the scalars: vector and double vector supermultiplets}
Due to the duality between scalar field, entering the action with
the space-time derivatives only, and gauge field strength in
$d=3$, the actions for the vector (one scalar dualized) and the
double vector (both scalars dualized) supermultiplets can be
easily obtained within the coset approach. Before performing such
dualizations, let us firstly rewrite our action \p{finAction} in
the vector notations. If we introduce the quantity \be\label{G}
\cG_{ab} =\frac{1}{\sqrt{2}}\left( \psi^m \cD_{ab} \bpsi_m -
\bpsi^m \cD_{ab} \psi_m\right), \ee then only vector indices show
up in the action. Passing to the vector notation, we will get
\bea\label{finAction_vec} S &=& 2\int d^3 x-
\int d^3 x \det \cE \left( 1+  \sqrt{(1-\cD_I q \cD_I\bar q)^2 -(\cD_{I} q \cD_I q)(\cD_{J}\bar q \cD_{J}\bar q)}\right)+ \nn \\
&& \im  \int d^3 x \;\det \cE \;\epsilon^{IJK} \cG_I \cD_{J}\; q\; \cD_K \; {\bar q} ,
\eea
where
\be
\cD_I \equiv \left( \cE^{-1}\right)_I{}^J \partial_J, \quad \cE_I{}^J= \delta_I^J-\frac{1}{\sqrt{2}}\left(\sigma^J\right)_{ab}
\left(\psi^a \partial_I \bpsi^b +\bpsi^a \partial_I \psi^b \right).
\ee
\subsection{Vector supermultiplet}
The standard $N=2,d=3$ supermultiplet included one scalar and one gauge fields (entering the action through the field strength) among the
physical bosonic components. Thus, we have to dualize one of the scalar fields in the action \p{finAction_vec}. To perform dualization,
firstly, one has to pass to the real bosonic fields $\{u, v\}$
\be
q=\frac{1}{2}(u+i v),\quad {\bar q} = \frac{1}{2}(u-i v).
\ee
In terms of newly defined scalars, the action \p{finAction_vec} reads
\bea\label{daction1}
S &=& 2\int d^3 x-
\int d^3 x \det \cE \left[ 1+  \sqrt{\left(1-\frac{1}{2}\cD_I u \cD_I u\right)\left(1-\frac{1}{2} \cD_J v \cD_J v\right) -\frac{1}{4} \left(\cD_{I} u \cD_I v\right)^2 }\right]+ \nn \\
&& \frac{1}{2}  \int d^3 x \;\det \cE \;\epsilon^{IJK} \cG_I \cD_{J}\; u\; \cD_K \; v ,
\eea
The equation of motion for bosonic field $v$ has the form
\be\label{deq1}
\partial_I \left( \det\cE \left(\cE^{-1}\right)_J^I\; V_J\right)=0,\quad V_I={\widetilde V}_I+\frac{1}{2}\epsilon_{IJK}G_J \cD_K u,
\ee
where
\be\label{deq2}
{\widetilde V}_I =\frac{\left(1-\frac{1}{2}\cD u \cdot \cD u\right) \cD_I v +\frac{1}{2}\ \cD u \cdot \cD v\; \cD_I u}
{2 \sqrt{\left(1-\frac{1}{2}\cD u \cdot \cD u\right)\left(1- \frac{1}{2}\cD v \cdot \cD v\right) -\frac{1}{4} \left(\cD u \cdot \cD v\right)^2 }}.
\ee
Then, one may find that
\be
\cD_I v = \frac{2 {\widetilde V}_I -{\widetilde V}\cdot \cD u \; \cD_I u}{\sqrt{1-\frac{1}{2} \cD u \cdot \cD u +2 {\widetilde V}\cdot {\widetilde V} -\left( {\widetilde V}\cdot \cD u\right)^2}} .
\ee
Now, performing the Rauth transformation over bosonic field $v$, we will finally get
\be\label{daction11}
{\tilde S}=2 \int d^3 x- \int d^3 x \det\cE \left( 1+\sqrt{1-\frac{1}{2} \cD u \cdot \cD u +2 {\widetilde V}\cdot {\widetilde V} -\left( {\widetilde V}\cdot \cD u\right)^2}\right).
\ee
This is the action for $N=2,d=3$ vector supermultiplet which possesses additional, spontaneously broken $N=2$ supersymmetry.

One should stress, that the real field strength is defined in
\p{deq1}, but the action has a much more simple structure being
written in terms of ${\widetilde V}_I$.

\subsection{Double vector supermultiplet}
Finally, one may dualize both scalars in the action \p{finAction_vec}. As the first step, one has to find the equations of motion for the scalar fields
\be\label{eomdv}
\partial_I \left( \det\cE \left( \cE^{-1}\right)^I_J\; V^J \right) =0, \quad \partial_I \left( \det\cE \left( \cE^{-1}\right)^I_J\; \bV^J \right) =0,
\ee where \be\label{VV} V_I ={\widetilde V}_I - \im \epsilon_{IJK}
G_J \cD_K {\bar q}, \quad {\widetilde V}_I= \frac{\left(1-\cD q
\cdot \cD {\bar q}\right) \cD_I {\bar q}+\left( \cD {\bar q} \cdot
\cD {\bar q}\right) \cD_I q}{\sqrt{(1-\cD q \cdot \cD \bar q)^2
-(\cD q \cdot \cD q)(\cD \bar q \cdot \cD\bar q)}}. \ee
After a
standard machinery with the Rauth transformations we will finally
get the action \be\label{dvector} {\widehat S}= 2\int d^3 x -\int
d^3 x \det\cE \left[1+ \sqrt{ \left( 1+ {\widetilde V}\cdot
{\overline{\widetilde V}}\right)^2 -{\widetilde V}{}^2\;
{\overline{\widetilde V}}{}^2} - \im \epsilon_{IJK}\; G_I\;
{\widetilde V}_J {\overline{\widetilde V}}_K\right] . \ee The
bosonic sector of this action coincides with that constructed in
\cite{IKL22}. Again, the simplest form of the action is achieved
with the help ${\widetilde V}_I$ variables which are related with
field strengths as in \p{eomdv}, \p{VV}.

\section{Conclusion}
In this paper, using a remarkable connection between partial
breaking of global supersymmetry, coset approach, which realized
the specific pattern of supersymmetry breaking, and the Nambu-Goto
actions for the extended object, we have constructed the on-shell
component action for $N=1, D=5$ supermembrane and for its dual
cousins. Of course, such an action can be obtained by dimensional
reduction from the superspace action of the 3-brane in $D=6$ (see
e.g., \cite{BG1}, \cite{RT}) or from the action of ref.\cite{HLP}.
Nevertheless, if we pay more attention to the spontaneously broken
supersymmetry and, thus, use the corresponding covariant
derivatives, together with the proper choice of the components,
the resulting action can be drastically simplified. So, the
implications of our results are threefold:
\begin{itemize}
\item we demonstrated that the coset approach can be used  far beyond the construction of the superfield equations of motion
if we are interested in the component actions,
\item we showed that there is a rather specific choice of the superfields and their components which drastically simplifies the component action,
\item we argued that the broken supersymmetry fixed the on-shell component action up to some constants, while the role
of the unbroken supersymmetry is just to fix these constants.
\end{itemize}
The application of our approach is not limited to the cases of P-branes only.
Different types of D-branes could be also considered in a similar
manner. However, once we are dealing with the field strengths,
which never show up as the coordinates of the coset space, the
proper choice of the components becomes very important. In
particular, the Born-Infeld-Nambu-Goto action \p{daction1}, we
constructed by the dualization of one scalar field, has a nice,
compact form in terms of the ``covariant'' field strength
${\widetilde V}_I$ which is related with the ``genuine'' field
strength, obeying the Bianchi identity, in a rather complicated
way \p{deq1}. The same is also true for the Born-Infeld type
action \p{dvector}. In order to clarify the nature of these variables, one
has to consider the corresponding patterns of the supersymmetry
breaking (with one, or without central charges in the $N=4,d=3$
Poincar\'{e} superalgebra \p{algebra}) independently. In this
respect, the detailed analysis of $N=2 \rightarrow N=1$
supersymmetry breaking in $d=4$ seems to be much more interesting,
being a preliminary step to the construction of $N=4$ Born-Infeld
action \cite{Kallosh, BIK22} and/or to the action describing
partial breaking of $N=1,D=10$ supersymmetry with the
hypermultiplet as the Goldstone superfield.

\section*{Acknowledgements}
We wish to acknowledge discussions with Anton Sutulin.

S.K. is grateful to INFN - Laboratori Nazionali di Frascati for
warm hospitality. This work was partially supported by RFBR
grants~12-02-00517-a, 13-02-91330-NNIO-a and 13-02-90602 Apm-a, as
well as by the ERC Advanced Grant no. 226455
\textit{``Supersymmetry, Quantum Gravity and Gauge
Fields''}~(\textit{SUPER\-FIELDS}).

\setcounter{equation}{0}
\def\theequation{A.\arabic{equation}}
\section*{Appendix A: Superalgebra, coset space, transformations and Cartan forms}
In this Appendix we collected some formulas describing the nonlinear realization
of $N=1, D=5$ Poincar\'{e} group in its coset over $d=3$ Lorentz group $SO(1,2)$.

In $d=3$ notation the $N=1, d=5$  Poincar\'e superalgebra contains the following set of generators:
\be\label{1}
\mbox{ N=4, d=3 SUSY }\quad \propto \quad \left\{ P_{ab}, Q_a,\bQ_a, S_a,\bS_a, Z, \bZ, M_{ab}, K_{ab}, \bK_{ab}, J \right\},
\ee
$a,b=1,2$ being the $d=3$ $SL(2,R)$
spinor indices \footnote{The indices are raised
and lowered as follows:
$V^{a}=\epsilon^{ab}V_b,\;V_{b}=\epsilon_{bc}V^c,\quad
\epsilon_{ab}\epsilon^{bc}=\delta_a^c\; .$}. Here, $P_{ab}$, $Z$  and $\bZ$ are $D=5$ translation generators,  $Q_a, \bQ_a$ and $S_a, \bS_a$ are the generators of super-translations, the generators $M_{ab}$ form $d=3$ Lorentz algebra $so(1,2)$, the generators $K_{ab}$ and $\bK_{ab}$ belong to the coset $SO(1,4)/SO(1,2)\times U(1)$, while $J$ span $u(1)$. The basic commutation relations read
\bea\label{algebra}
&&\left[ M_{ab}, M_{cd}  \right]=\epsilon_{ad}M_{bc}+\epsilon_{ac}M_{bd}+\epsilon_{bc}M_{ad}+\epsilon_{bd}M_{ac} \equiv \left( M\right)_{ab,cd},  \nn \\
&&\left[ M_{ab}, P_{cd}  \right]=\left( P\right)_{ab,cd}, \;
\left[ M_{ab}, K_{cd}  \right]=\left( K\right)_{ab,cd}, \;
\left[ M_{ab}, \bK_{cd}  \right]=\left( \bK\right)_{ab,cd},  \nn \\
&&\left[ K_{ab}, \bK_{cd}  \right] = \frac{1}{2}\left( M\right)_{ab,cd} +2\left( \epsilon_{ac}\epsilon_{bd}+  \epsilon_{bc}\epsilon_{ad}  \right)J, \nn \\
&&\left[ K_{ab}, P_{cd}   \right] = -\left( \epsilon_{ac}\epsilon_{bd}+  \epsilon_{bc}\epsilon_{ad}  \right) Z, \ \left[ \bK_{ab}, P_{cd}   \right] = \left( \epsilon_{ac}\epsilon_{bd}+  \epsilon_{bc}\epsilon_{ad}  \right) \bZ,  \nn \\
&& \left[ K_{ab}, \bZ   \right]  = -2P_{ab}, \ \left[ \bK_{ab}, Z   \right]  = 2P_{ab},\;
\left[ M_{ab}, Q_{c}  \right] = \epsilon_{ac}Q_{b}+\epsilon_{bc}Q_{a}\equiv \left( Q\right)_{ab,c},
\nn\\
&&  \ \left[ M_{ab}, \bQ_{c}  \right] = \left( \bQ\right)_{ab,c},  \;
 \left[ M_{ab}, S_{c}  \right] = \left( S\right)_{ab,c}, \ \left[ M_{ab}, \bS_{c}  \right] = \left( \bS\right)_{ab,c},  \nn \\
&& \left[ \bK_{ab}, Q_{c}  \right] =- \left( \bS\right)_{ab,c}, \ \left[ K_{ab}, \bQ_{c}  \right] = \left( S\right)_{ab,c},  \;
\left[ \bK_{ab}, S_{c}  \right] = \left( \bQ\right)_{ab,c}, \ \left[ K_{ab}, \bS_{c}  \right] = -\left( Q\right)_{ab,c},  \nn \\
&& \left[ J,Q_{a}\right]=-\frac{1}{2}Q_a,\;\left[ J,\bQ_{a}\right]=\frac{1}{2}\bQ_a,\;
\left[ J,S_{a}\right]=-\frac{1}{2}S_a,\;\left[ J,\bS_{a}\right]=\frac{1}{2}\bS_a,\nn \\
&& \left[ J,K_{ab}\right]=-K_{ab},\;\left[ J,\bK_{ab}\right]=\bK_{ab},\;
\left[ J,Z\right]=-Z,\;\left[ J,\bZ\right]=\bZ,\nn \\
&& \left\{ Q_{a} , \bQ_{b}  \right\} =2P_{ab}, \ \left\{ S_a, \bS_b \right\}=2P_{ab},  \; \left\{ Q_{a}, S_{b} \right\} =2\epsilon_{ab} Z, \ \left\{ \bQ_{a}, \bS_{b} \right\} =2\epsilon_{ab} \bZ.
\eea
Note, that the generators obey the following conjugation rules:
\bea\label{conrules}
&& \left(P_{ab}\right)^\dagger=P_{ab},\;\left(K_{ab}\right)^\dagger=\bK_{ab},\;\left(M_{ab}\right)^\dagger=-M_{ab},\;
J^\dagger=J,\;Z^\dagger=\bZ,\nn \\
&& \left(Q_{a}\right)^\dagger=\bQ_{a},\;\left(S_{a}\right)^\dagger=\bS_{a}.
\eea
We  define the coset element as follows
\be\label{coset}
g = e^{\im x^{ab}P_{ab}}e^{\theta^a Q_a + \bar\theta^a \bQ_a}e^{\im(\mq Z+\bar\mq \bZ)}e^{\mpsi^a S_a + \mbpsi^a \bS_a} e^{\im  (\mLambda^{ab}K_{ab}+ \mbLambda^{ab}\bK_{ab})}.
\ee
Here, $\left\{x^{ab}, \theta^a, \bar\theta^a\right\}$ are $N=2, d=3$ superspace coordinates, while the remaining coset parameters are Goldstone superfields,
$ \mpsi^a \equiv \mpsi^a(x,\theta, \bar\theta),\;\mbpsi^a \equiv \mbpsi^a(x,\theta, \bar\theta),\;\mq \equiv \mq(x,\theta, \bar\theta),\;{\bar\mq} \equiv {\bar\mq}(x,\theta, \bar\theta),\; \mLambda^{ab}
\equiv \mLambda^{ab}(x,\theta, \bar\theta),\; \mbLambda^{ab}
\equiv \mbLambda^{ab}(x,\theta, \bar\theta)$. These $N=2$ superfields obey the following conjugation rules
\be\label{conrules1}
\left( x^{ab}\right)^\dagger=x^{ab},\; \left( \theta^a\right)^\dagger =\bar\theta{}^a,\quad
\mq^\dagger=\bar\mq,\; \left( \mpsi^a\right)^\dagger =\mbpsi^a,\; \left(\mLambda^{ab}\right)^\dagger =\mbLambda^{ab}.
\ee
The transformation properties of the coordinates and superfields with respect
to all symmetries can be found by acting from the left on the coset element $g$ \p{coset} by the different elements of $N=1, D=5$ Poincar\'{e} supergroup. In what follows, we will need the explicit form only for the broken $(S, \bS)$ and unbroken $(Q,\bQ)$ supersymmetries, and $(K, \bK)$ automorphism transformations which read:
\begin{itemize}
\item Unbroken $(Q)$ supersymmetry $(g_0=\mbox{exp }(  \epsilon^{a}Q_{a}+\bar\epsilon^{a}\bQ_{a} ))$
\be\label{susy1}
\delta x^{ab}=\im \left( \epsilon^{(a}\bar\theta^{b)}+\bar\epsilon^{(a}\theta^{b)}\right) ,
\quad
\delta \theta^{a}=\epsilon^a\;,\delta \bar\theta^{a}=\bar\epsilon^a\; .
\ee
\item Broken $(S)$ supersymmetry $(g_0=\mbox{exp }\left(  \varepsilon^{a}S_{a}+\bar\varepsilon^{a}\bS_{a} \right))$
\be\label{susy2}
\delta x^{ab}= \im \left(\varepsilon^{(a}\mbpsi^{b)}+\bar\varepsilon^{(a}\mpsi^{b)}\right),\quad
\delta \mq=2\im \varepsilon_a\theta^a,\;\delta \bar\mq=2\im \bar\varepsilon_a\bar\theta^a,\quad
\delta\mpsi^a=\varepsilon^a,\;\delta\mbpsi^a=\bar\varepsilon^a\; .
\ee
\item Automorphism $(K,\bK)$ transformations  $(g_0=\mbox{exp }\im\left(  a^{ab}K_{ab}+{\bar a}{}^{ab}\bK_{ab} \right))$
\bea\label{auto}
&& \delta x^{ab}= -2\im \left(a^{ab}\mq-{\bar a}{}^{ab}{\bar\mq}\right)-2\theta^c\mpsi_c {\bar a}{}^{ab}+
2\bar\theta{}^c\mbpsi_c a^{ab},\quad \delta\theta^a=-2\im a^{ab}\mbpsi_b,\;\delta\bar\theta^a=2\im {\bar a}^{ab}\mpsi_b,  \nn \\
&& \delta \mq=- 2\im a^{ab}x_{ab}-2 a^{ab}\left( \theta_a \bar\theta_b - \mpsi_a\mbpsi_b\right),\quad
\delta\mpsi^a=2 \im a^{ab}\bar\theta_b, \nn \\
&& \delta \bar\mq= 2\im \bar{a}^{ab}x_{ab}-2 {\bar a}^{ab}\left( \theta_a \bar\theta_b - \mpsi_a\mbpsi_b\right),\quad
\delta\mbpsi^a=-2\im {\bar a}^{ab}\theta_b.
\eea
\end{itemize}

As the next step of the coset formalism, one construct the Cartan forms
\be\label{CFdef}
g^{-1}d g =\Omega_P+\Omega_Q+\overline\Omega_Q+\Omega_Z+\overline\Omega_Z+\Omega_S+\overline\Omega_S+\ldots\;.
\ee
In what follows we will need only the forms $\Omega_P, \Omega_{Q},  \Omega_{Z}$ and $\Omega_{S}$ which explicitly read
\bea
&&\Omega_P =\left\{ \left(\cosh{2\sqrt{\mY}}\right)_{ab}^{cd}\triangle x^{ab} -\im \left( \mbLambda^{ab}\triangle \mq - \mLambda^{ab}\triangle \bar\mq \right)\left( \frac{\sinh{2\sqrt{\mY}}}{\sqrt{\mY}} \right)_{ab}^{cd} \right\} P_{cd}, \nn \\
&&\Omega_{Q} =\left\{d \theta^{b}  \left(\cos{2\sqrt{\mbT}}\right)_{b}^{\;\, c}- \im\, d \mbpsi^{b} \,\mLambda_{b}^{\;\, a} \left(\frac{\sin{2\sqrt{\mbT}}}{\sqrt{ \mbT}}\right)_{a}^{\;\, c}\right\} Q_c,\nn\\
&& \Omega_Z = \left\{  \triangle \mq  +  \left( \mbLambda^{ab} \triangle \mq - \mLambda^{ab}\triangle\bar\mq \right) \left( \frac{\cosh 2\sqrt{\mY} -1}{\mY} \right) _{ab}^{cd} \mLambda_{cd} + \im dx^{ab} \left( \frac{\sinh{2\sqrt{\mY}}}{\sqrt{\mY}}  \right)_{ab}^{cd} \mLambda_{cd}  \right\}Z, \nn \\
&&\Omega_{S} =\left\{d \mpsi^{b}  \left(\cos{2\sqrt{\mbT}}\right)_{b}^{\;\, c}+ \im\, d \bar\theta^{b} \,\mLambda_{b}^{\;\, a} \left(\frac{\sin{2\sqrt{\mbT}}}{\sqrt{ \mbT}}\right)_{a}^{\;\, c}\right\} S_c,\label{CF}\\
&& \triangle x^{ab} = dx^{ab} -\im \left( \theta^{(a} d\btheta^{b)} + \btheta^{(a} d\theta^{b)}+
\mpsi^{(a} d\mbpsi^{b)} + \mbpsi^{(a} d\mpsi^{b)}\right), \label{Dx}\\
&&\triangle \mq = d\mq -2\im \mpsi_{a}d\theta^{a}, \; \triangle \bar\mq = d\bar\mq -2\im \mbpsi_{a}d\bar\theta^{a} .\label{Dq}
\eea
Here, we defined matrix-valued functions $\mY_{ab}{}^{cd}, \mT_a{}^b$ and ${\mbT}{}_a{}^b$ as
\be\label{YT}
\mY_{ab}{}^{cd} = \mLambda_{ab}\mbLambda^{cd}+ \mbLambda_{ab}\mLambda^{cd},\quad
\mT_a{}^b=\mLambda_{a}{}^c\mbLambda_{c}{}^b, \;{\mbT}_a{}^b=\mbLambda_{a}{}^c\mLambda_{c}{}^b .
\ee
Note, that all these Cartan forms transform homogeneously under all symmetries.

Having at hands the Cartan forms, one may construct the ``semi-covariant'' (covariant with respect to $d=3$ Lorentz,
unbroken and broken supersymmetries only) derivatives as
\be\label{cD}
{\triangle x}^{ab}\nabla_{ab} +d\theta^a \nabla_a+d\bar\theta^a \bar\nabla_a = dx^{ab} \frac{\partial}{\partial x^{ab}} + d\theta^a \frac{\partial}{\partial \theta^a}+d\bar\theta^a \frac{\partial}{\partial \bar\theta^a}.
\ee
Explicitly, they read
\bea\label{nabla}
&& \nabla_{ab} = (E^{-1})_{ab}{} ^{cd} \partial_{cd},\nn \\
&&\nabla_a = D_a -\im \left( \mpsi^b D_a \mbpsi^c + \mbpsi^b D_a \mpsi^c  \right)\nabla_{bc} = D_a-\im \left( \mpsi^b \nabla_a \mbpsi^c + \mbpsi^b \nabla_a \mpsi^c  \right)\partial_{bc},
\eea
where
\bea
&& D_a=\frac{\partial}{\partial \theta^a} -\im\,\bar \theta^b\,\partial_{ab},\;
\bD_a=\frac{\partial}{\partial \bar\theta^a} -\im\, \theta^b\,\partial_{ab},\quad
\left\{ D_a, \bD_b\right\} = -2 \im \partial_{ab}, \label{flatCD} \\
&& E_{ab}{}^{cd} = \delta^{(c}_a \delta^{d)}_b -\im \left( \mpsi^{(c}\partial_{ab} \mbpsi^{d)} +\mbpsi^{(c}\partial_{ab} \mpsi^{d)}   \right),\label{E} \\
&& (E^{-1})_{ab}{}^{cd}=\delta^{(c}_a \delta^{d)}_b + \im \left( \mpsi^{(c}\nabla_{ab} \mbpsi^{d)} +\mbpsi^{(c}\nabla_{ab} \mpsi^{d)}   \right) \label{Em1}.
\eea
These derivatives obey the following algebra:
\bea\label{deralg}
&&\left\{\nabla_a, \nabla_b\right\}= -2\im\, \left( \nabla_a \mpsi^c  \nabla_b \mbpsi^d + \nabla_a \mbpsi^c  \nabla_b \mpsi^d  \right)\nabla_{cd} , \nn \\
&&\left\{\nabla_a, \bnabla_b\right\}=-2\im\, \nabla_{ab} -2\im \left( \nabla_a \mpsi^c  \bnabla_b \mbpsi^d + \nabla_a \mbpsi^c  \bnabla_b \mpsi^d  \right)\nabla_{cd} , \nn \\
&&\left[\nabla_{ab},\nabla_c\right] = -2\im\, \left( \nabla_{ab} \mpsi^d  \nabla_c \mbpsi^f + \nabla_{ab} \mbpsi^d  \nabla_c \mpsi^f  \right)\nabla_{df} ,\nn\\
&& \left[\nabla_{ab}, \nabla_{cd}\right] =2\im \left(   \nabla_{ab} \mpsi^m \nabla_{cd} \mbpsi^n -  \nabla_{cd} \mpsi^m \nabla_{ab} \mbpsi^n      \right) \nabla_{mn}.
\eea

To complete this rather technical Appendix, we will also define the  $d=3$ volume form  in a standard manner as
\be\label{volume}
d^3 x \equiv \epsilon_{IJK} dx^I \wedge dx^J \wedge dx^K \quad \Rightarrow \quad
dx^I \wedge dx^J \wedge dx^K = \frac{1}{6} \epsilon^{IJK} d^3 x.
\ee
The translation to the vectors is defined as
\be\label{44}
V^I \equiv \frac{\im}{\sqrt{2}}\left(\sigma^I\right)_a{}^b\; V_b{}^a \quad \Rightarrow \quad V_a{}^b =-\frac{\im}{\sqrt{2}} V^I \left( \sigma^I\right)_a{}^b,  \qquad V^{ab}V_{ab} = V^I V^I.
\ee
Here we are using the standard set of $\sigma^I$ matrices
\be\label{41}
\sigma^I \; \sigma^J = \im \epsilon^{IJK} \sigma^K + \eta^{IJ} E, \quad
\left( \sigma^I\right)_a{}^b \; \left(\sigma^I \right)_c{}^d = 2 \delta_a{}^d \delta_c{}^b - \delta_a{}^b \delta_c{}^d,
\ee
were $\epsilon^{IJK}$ obeys relations
\be\label{42}
\epsilon^{IJK}\epsilon_{IMN} =\delta^{J}_{M}\delta^{K}_{N} -\delta^{J}_{N}\delta^{K}_{M},\quad \epsilon^{IJK}\epsilon_{IJN}=2 \delta^{K}_{N}, \quad \epsilon^{IJK}\epsilon_{IJK}=6.
\ee

\end{document}